
\magnification=1200
\baselineskip=21pt
\overfullrule=0pt
\null
\vskip1truein
\centerline{\bf GREEN FUNCTIONS}
\centerline{\bf AND}
\centerline{\bf THERMAL NATURE OF BLACK-HOLES}
\vskip0.5truein
\centerline{Milan Miji\'c\footnote*{milan@moumee.calstatela.edu}}
\bigskip
\centerline{\it Department of Physics and Astronomy}
\centerline{\it California State University, Los Angeles, CA 90032}
\vskip1truein
ABSTRACT:
There are several ways to establish and study thermal properties of
black holes. I review here method of Fulling and Ruijsenaars, based
on the analytic structure of Green functions on the complex plane.
This method provides a clear distinction between zero and finite temperature
field theories, and allows for quick evaluation of black hole temperature.

\vskip0.25truein
\centerline{\sl Two lectures at 3rd Danube Workshop, June 1993,
Belgrade, Yugoslavia}
\vfil\eject

{\bf Introduction.}
\bigskip
Quantum evaporation of black holes exhibits fascinating connection between
gravity, quantum mechanics and thermodynamics. In order to understand this
phenomena better, and certainly in order to move on towards solution of
bigger questions, like the back-reaction, the issue of the final state, or
the possible evolution from pure to mixed state, (otherwise known as the
\lq\lq loss of quantum coherence\rq\rq), it would be usefull to examine
different paradigmas and different techniques that have been used in the
past for study of black hole evaporation. Historically the first has been
Hawking's use of mode mixing and Bogoliubov coefficients [1]. Close to
this is a technique utilizing properties of the squeezed states in black hole
spacetimes, which received some attention recently [2]. Next, there
is a puzzling and not yet fully understood discovery of a relationship
between the flux of the outgoing radiation and the trace anomaly [3],
which has been used most recently
in studies of dilaton models [4]. Another approach explored in
seventies has been in the physical analogy with the Klein paradox [5].
But most people consider as the most superior the method
of thermal Green functions, introduced by Gibbons and Perry [6]. This is
just the method that we will explore here, but our starting point will be
a comprehensive study of the analytic properties of Green
functions, done by Fulling and Ruijsenaars [7]. We will pragmatically follow
their path, by computing first a number of relevant quantities for scalar field
in flat spacetime at zero and at the finite temperature. Next we move on to
curved spacetime and discuss thermal properties of Rindler and Schwarzschild
spacetime. This methods allows for easy and clean evaluation of the
temperature of the black hole as seen by the stationary observer outside
the horizon.

\vfil\eject

{\bf Scalar field in flat spacetime}
\bigskip

The basic relations for a massive scalar field are
$$
S[\Phi] = \int d^4x~\left [ {1 \over 2} \partial \Phi \cdot \partial \Phi +
{1 \over 2} m^2 \Phi^2 \right ]~~;
$$

$$
\pi = \dot {\Phi}~~;~~~~~~~~~~~~~~
(- \partial^2 + m^2 ) \Phi (x) = 0~~;
$$

$$
\Phi (x) = \int d\vec{k}~\left [ a_{\vec k} \phi_{\vec k}(x) +
a_{\vec k}^{\dag}\phi_{\vec k}^*(x) \right ]~~;~~~~~~~~
[a_{\vec k}, a_{\vec k}^{\dag}] = \delta (\vec {k} - \vec {q})~~.
$$

$$
\phi_{\vec k}(x) = {e^{i k \cdot x} \over 2 \omega (k)}~~;~~~~~~~~~~~
\omega (k) \equiv + \sqrt {\vec {k}^2 + m^2}~~.
$$

$$
[\Phi (t, \vec {x}), \Phi (t, \vec {y})] = 0~~;~~~~~~~~
[\Phi (t, \vec {x}), \pi (t, \vec {y})] = i \delta (\vec {x} - \vec {y})~~.
$$

\bigskip
Suppose now that we want to compute the two point corellation
function
$$
\langle \Phi (x) \Phi (y) \rangle ~~.
\eqno (1)$$
The two interesting choices for the quantum state in which this
expectation value is to be computed are vacuum and the thermal state. We
will consider them now in turn.

\vskip0.5truein

{\bf Vacuum expectation values}

\bigskip
If we choose to compute expectation values in the vacuum state, it follows
that,
$$
\langle 0| \Phi (x), \Phi (y) |0 \rangle =
\int d\vec {k}~{e^{i \vec {k} \cdot (\vec {x} - \vec {y}) - i k^0 (x^0 - y^0)}
\over 2 \omega (k)}~~.
$$

Using Cauchy formula,
$$
f(z_0) = {1 \over 2 \pi i} \int_{C(z_0)} dz {f(z) \over z - z_0} ~~,
$$
we can rewrite this as,
$$
\langle 0 | \Phi (x) \Phi (y) | 0 \rangle =
(-i) \int {d^4k \over (2\pi)^4} ~{e^{-ik\cdot \Delta x} \over (k^0)^2 -
\omega(k)^2} ~~.
\eqno (2)$$

The correspodence between these two expressions is not unique, as the later
integral is multivalued. The integrand has two poles, at
$k^0 = \pm \omega (k)$, and the final result depends on how does the contour
of integration go around the poles. This choice of contour is
equivalent to the choice of boundary condition for the corellation function.
The residua at the
two poles are $(-1)/(2 k_{pole}^0)$, and the two basic contributions are
$$
{- i \over 2 \omega (k)} e^{-i\omega (k) t}~~,~~~~~{\rm and}~~~~~~
{i \over 2 \omega (k)} e^{i \omega (k) t} ~~.
$$
So, the general result will be,
$$
\langle 0| \Phi (x), \Phi (y) |0 \rangle_C =
\int {d\vec {k} \over (2\pi)^3} ~ e^{i \vec {k} \cdot \vec {x}}~
\sum_C~(- i) {e^{k_{pole}^0 x^0} \over 2 k_{pole}^0} ~~,
$$
where $C$ stands for the contour of integration.
The 2-point correlation function in Eq. (2) corresponds therefore to one
particular choice for the contour of integration in (2), or to one particular
choice of the boundary conditions for the solutions of Eq.'s (3-4) below.
Different choice of contour leads to different vacuum expectation value in
Eq. (1).

Different vacuum expectation values have been historically
defined as follows:

$$
G^+(x, y) \equiv \theta (x^0 - y^0)
\langle 0| \Phi (x) \Phi (y) | 0 \rangle
{}~~~~ {\rm (positive
{}~frequency~Wightman~function)};
$$

$$
G^-(x, y) \equiv \theta (y^0 - x^0)
\langle 0| \Phi (y) \Phi (x) | 0 \rangle ~~~~ {\rm (negative
{}~frequency~Wightman~function)};
$$
\medskip
$$
iG(x, y) \equiv \langle 0| [\Phi (x), \Phi (y)] | 0 \rangle ~~~~{\rm
(Pauli-Jordan
{}~function)};
$$

$$
G^{(1)}(x, y) \equiv \langle 0| \{\Phi (x), \Phi (y)\} | 0 \rangle ~~~~
{\rm (Hadamard~function)};
$$
\medskip
$$
G_R(x, y) \equiv (-) \theta (t_x - t_y)~G(x, y) ~~~~ {\rm (retarded~Green
{}~function)};
$$

$$
G_A(x, y) \equiv \theta (t_y - t_x)~G(x, y)~~~~ {\rm
(advanced~Green~function)};
$$
\medskip
$$
iG_F(x, y) \equiv \langle 0| T(\Phi (x) \Phi (y)) | 0 \rangle ~~~~ {\rm
(Feynman
{}~propagator)};
$$

\bigskip
All of these 2-point functions are usually called Green functions although
four of them actually obey the homogeneous equation. There is a number of
useful and easy to prove relations that are obeyed by these functions:

{\it Problem 1.} Show that,
$$
iG = G^+ - G^-~~;~~~~~~~~
G^{(1)} = G^+ + G^- ~~;~~~~~~~
i G_F = G^+ + G^- ~~.
$$

{\it Problem 2.}. Show that,
$$
(- \partial_x^2 + m^2)~{\cal G}(x, y) = 0 ~~,
\eqno (3)$$
when ${\cal G} \in \{G^+, G^-, iG, G^{(1)} \}$.

{\it Problem 3.} {\it (a)} Show first that,
$\partial_{tt}^2 (\theta G) = \delta \dot {G} + \theta \ddot {G}$; then,
{\it (b)}
show that the other three vacuum expectation values are true Green functions:

$$
(- \partial_x^2 + m^2)~{\cal G}(x, y) = \pm \delta (x - y) ~~,
\eqno (4)$$
when ${\cal G} \in \{G_R, G_A, G_F \}$. Plus sign applies for the first
two, minus sign for $G_F$.

Further claim is that solution to any of these equations has the following
integral representation:

$$
{\cal G}(x, y) = \int {d^4k \over (2\pi)^4} ~e^{ik\cdot (x - y)}~{\cal G}(k)
{}~~,
\eqno (5)$$
which obeys,

$$
(-\partial_x^2 + m^2) {\cal G}(x, y) = \int {d^4 k \over (2\pi)^4}~
e^{ik\cdot (x-y)}~
(k^2 + m^2){\cal G}(k)~~.
$$

For ${\cal G} \in \{G^+, G^-, iG, G^{(1)} \}$, the right hand side of this
equation should vanish; when ${\cal G} \in \{G_R, G_A, G_F \}$, the integrand
of the Fourier transform on the right hand side should be $\pm 1$. Both
can be accomplished using the multivaluedness of the complex integral (5),
as the following excersizes will show:

{\it Problem 4.} Draw all seven contours of integration for
defined Green functions, and discuss boundary conditions.

{\it Problem 5.}
Since first four functions obey the same equation, they are distinguished
by different boundary conditions, and similarly for the other three:

{\it (a)} Determine boundary conditions at $\pm \infty$ for four
solutions to the homogeneous equation, and justify names and notation for
$G^{\pm}$.

{\it (b)} Determine boundary conditions for the three solutions to the
inhomogeneous equation (4)
by adding a source term $J(x) \Phi (x)$ to the original Lagrangean.

Let us now fix $\vec x - \vec y \rightarrow \vec x$ as a parameter, and
consider all these Green functions as functions of time. It is immediately
obvious that, by virtue of its defining expression, $G^+(t)$ may be
analytically extended only to the upper half-plane, and $G^-(t)$ only to the
lower. Further straigthforward examination uncovers the following beautifull
structure on the complex plane $z \equiv (t, \tau)$:

{\it Problem 6.} Show that there is an analytic function ${\cal G}(z)$ which
coincides with the $G^+$ on the upper half-plane, and with $G^-$ on the lower
half-plane. Along the $t$ axis ${\cal G}$ has a branch cut for $t > |\vec x|$.
The jump accross the cut is given by the Pauli-Jordan Green function:
${\cal G}(t + i\epsilon) - {\cal G}(t - i\epsilon) = G(t)$, for $\epsilon > 0$.

{\it Problem 7.} The projection of {\cal G} to the imaginary axis is called the
Euclidean, or the Schwinger Green function: ${\cal G}(z) \rightarrow_{t
\rightarrow 0} G_E(\tau)$. Show that:

{\it (a)} $G_E$ obeys the Euclidean version of the inhomogeneous field equation
for a Green function,
$$
(\partial_{\tau \tau}^2 + \nabla^2 + m^2)G_E(\tau, \vec {x}) = - \delta
(\tau) \delta^{(3)}(\vec {x}) ~~;
$$

{\it (b)} $G_E$ and $G_F$ are related by  the so-called Wick rotation:
$t \rightarrow i\tau$.

\bigskip
Results of these excersize may be stated as follows.
There is Green function ${\cal G}(z)$, which in
upper half-plane represents the analytic continuation of $G^+(t)$, and in
lower half-plane it is the analytic continuation of $G^-(t)$. Along the
real axis ${\cal G}$ has a branch cut for $t \geq |\vec x|$. The jump accross
the branch cut is given by the Pauli-Jordan Green function, that is, by the
canonical commutator for the scalar field $\phi$. On the imaginary axis
${\cal G}$ reduces to the Euclidean, or Schwinger Green function $G_E(\tau)$.
Unlike all the others mentioned so far, $G_E$ is a true Green function: it
obeys the Euclidean version of the field equation with the point source.
$G_E(\tau)$ by itself has its own analytic continuation to the real axis:
its is the Feynman propagator, $G_F(t)$, which obeys the inhomogeneous
field equation with the point source. This continuation is usually called
the Wick rotation.

This description, first given so eloquently by Fulling and
Ruijsenaars, is the culmination of our understanding of the Green functions.

\bigskip
We can now compute explicitly these functions. The most efficient strategy
is to  compute $G^{\pm}$ first, by evaluating the integral
above with the appropriate boundary conditions. Other Green functions may be
then computed from $G^{\pm}$, by adding or subtracting the two residua
according
to their contours, and/or the algebraic relations between the functions.
Results of the excersizes below should be compared to the calculations
in textbook of Bogoliubov and Shirkov [8].

{\it Problem 8.} Compute $G^+$ for a massless scalar field in flat spacetime.
Since there is no free parameter in the Lagrangean there is no characteristic
correlation scale and this Green function describes a long range correlations.

{\it Problem 9.} Compute $G^+$ for a massive scalar field in a flat spacetime.
This Green function does have a characteristic correlation length, given by
the finite mass of the particle. In the perturbation theory this would lead to
the Yukawa potential.

{\it Problem 10.} Now compute all the other Green functions for a massless
scalar field in flat spacetime.

{\it Problem 11.} What can we conclude from these Green functions about the
physics of a massless scalar field?

{\it Problem 12.} Compute all the Green functions for a massive scalar field
in flat spacetime.

{\it Problem 13.} What can we conclude from these Green functions about the
physics of a massive scalar field?

\vskip0.5truein

{\bf Thermal expectation values}

\bigskip

Instead of the vacuum, we now consider the thermal state,
$$
\rho = \sum_E |E\rangle e^{-\beta E} \langle E| ~~.
$$
The 2-point correlation function is,
$$
\langle \Phi(x) \Phi(y) \rangle = \sum_E e^{-\beta E} \langle E | \Phi (x) \Phi
(y)
|E\rangle~~.
$$
It is straightfoward to show that the later can be written in the path-integral
form. For this we should imagine that $\beta$ is the total interval in the
imaginary time within which the \lq\lq evolution \rq\rq takes place:
$\beta = \tau_f - \tau_i$. Separating this interval into the infinitesimal
ones, and inserting the unit operators represented through complete sums
of the coordinate or momentum eigenstates will lead to the formal
expression,
$$
\langle out|\Phi(\vec x, \tau_x) \Phi(\vec y, \tau_y) | in \rangle =
\int_{{\cal C}(\Phi)}\left [d\Phi(\vec r, \tau)\right ]
{}~e^{-S_E[\Phi(\vec r, \tau)]}
\Phi(\vec x, \tau_x)
\Phi(\vec y, \tau_y) ~~.
$$
Here ${\cal C}(\Phi)$ stands for the class of the fields over which the
path integral is taken. In this case they are those with the
periodic boundary condition in the imaginary time:
$$
{\cal C}(\Phi) = \{\Phi (\vec r, \tau)~~ |~~ \Phi (\vec r, \tau_i + \beta) =
\Phi (\vec r, \tau_i); \tau_i \leq \tau \leq \tau_f = \tau_i + \beta \}~~.
$$
This kind of reasoning is widely used, but it usually leaves some ambiguity
as to what is exactly the relationship between the fields and the Physics in
the real and imaginary times. Practice has shown what is the safe
way to interpret this relation in many cases, but the situation often gets
ambiguous when the curved spacetime is introduced.
It is in this methodological respect that the construction
of Fulling and Ruijsenaars appears to us so effective. We will now repeat the
preceding analysis from the vacuum case to determine the analytic structure
of the Green functions in the case of a free field in the flat spacetime,
but at the finite temperature $T \equiv \beta^{-1}$.

\bigskip
The two Whightman functions are defined as follows:
$$
G_{\beta}^+ (t, \vec x, \vec y)
\equiv Z^{-1}\sum_n Tr\{e^{-\beta H} \Phi(t, \vec x)
\Phi^{\dagger}(0, \vec y) \}~~;
$$
$$
G_{\beta}^- (t, \vec x, \vec y)
\equiv Z^{-1}\sum_n Tr\{e^{-\beta H}
\Phi^{\dagger}(0, \vec y)
\Phi(t, \vec x)
\}~~.
$$
The usual interpretation is that $G_{\beta}^+$ is an amplitude to create
particle at $(0, \vec y)$ and annihilate it at $(t, \vec x)$, all in the
a thermal bath (rather than vacuum).

Let $N_n^+$ denote the number of particles with positive energy $\omega_n >0$,
and $N_n^-$ number of antiparticles with the same energy.
The partition function is given as,
$$
Z = \sum_{\{N_n^+, N_n^-\}}
e^{-\beta \sum_n(N_n^+ + N_n^-)\omega_n}=
\prod_n \left (1 - e^{-\beta\omega_n} \right )^{-2}~~.
$$
The power of two appears here if we are dealing with a complex, non-hermitean
field. If $\Phi = \Phi^{\dagger}$, then,
$$
Z = \prod_n \left (1 - e^{-\beta \omega_n} \right )^{-1}~~.
$$

As in the vacuum case, where $\langle 0| a a |0\rangle =
\langle 0| a^{\dagger}a^{\dagger} |0\rangle = 0$, we have
$Tr\{\rho a a\} = Tr\{\rho a^{\dagger}a^{\dagger} \} =0$. However,
instead of $\langle 0| a^{\dagger} a|0\rangle =0$, we have the finite
occupation numbers for thermal state:
$$
\langle a_n a_m^{\dagger} \rangle = \delta_{nm}
\left (1 - \omega_n^{-1}{\partial \over \partial \beta} \log Z \right)=
\delta_{nm} \left (1 - e^{-\beta \omega_n} \right)^{-1}~~,
$$
so, that,
$$
\langle a_m^{\dagger}a_n \rangle = - 1 + \langle a_n a_m^{\dagger} \rangle =
\left ( e^{\beta \omega_n} - 1 \right )^{-1} ~~.
$$

Using the expansion for the quantum field with time dependent piece explicitly
written out (the spatial piece depends on the choice of the spatial coordinates
and topology),
$$
\Phi (t, \vec x) = \sum {f_n(\vec x) \over \sqrt{2 \omega_n}}
\left ( a_n e^{-i\omega_n t} + a_n^{\dagger} e^{i \omega_n t} \right )~~,
$$
we find the following expressions for the Whightman functions:
$$
G_{\beta}^+(t; \vec x, \vec y) =
\sum_n {f_n(\vec x)f_n^*(\vec y) \over 2\omega_n}
\left [
{e^{-i\omega_n t} \over 1 - e^{-\beta \omega_n}} +
{e^{i\omega_n t} \over  e^{\beta \omega_n} - 1}
\right ]~~;
$$
$$
G_{\beta}^-(t; \vec x, \vec y) =
\sum_n {f_n(\vec x)f_n^*(\vec y) \over 2\omega_n}
\left [
{e^{-i\omega_n t} \over e^{\beta \omega_n} - 1} +
{e^{i\omega_n t} \over  1 - e^{- \beta \omega_n}}
\right ]~~.
$$

\medskip
Let us now look at the analytic coninuation of these functions. The first
step is to establish their relationship to the Pauli-Jordan function:

{\it Problem 14.} Check that still,
$$
G_{\beta}^+ - G_{\beta}^- = [ \Phi(t, \vec x), \Phi (t, \vec y) ]~~.
$$

Since exponents with both signs are present in the expressions for
$G_{\beta}^{\pm}$, if we now substitute $z = t + i\tau$ for $t$, we can see
that neither of the function is analytic on the whole half-plane, either
upper or lower. However, observe that if we fix $\tau >0$, for
$\omega_n \rightarrow \infty$,
$$
G_{\beta}^+ \sim e^{\omega_n \tau}~~, ~~~~~G_{\beta}^- \sim e^{\omega_n
(\tau - \beta)}~~.
$$

Similarly, for $\tau < 0$, the leading pieces are,
$$
G_{\beta}^+ \sim e^{-\omega_n(\tau + \beta)}~~, ~~~~~
G_{\beta}^- \sim e^{-\omega_n \tau}~~.
$$

Therefore, $G_{\beta}^+$ may be extended from the real axis to the strip
$\tau \in [-\beta, 0]$, and $G_{\beta}^-$ to the strip $\tau \in [0, \beta]$.

Furthermore:

{\it Problem 15.} Observe that,
$$
G_{\beta}^+(\tau - \beta) = G_{\beta}^- (\tau)~~;~~~~~~~~~~~~
G_{\beta}^-(\tau + \beta) = G_{\beta}^+ (\tau)~~.
$$

Using this result it is easy to see that one can do analytic continuation
to the whole plane, using strips of the width $\beta$ in imaginary direction
as Weierstrass circles. One therefore arrives on the following analytic
structure on the complex plane: there is a complex function
${\cal G}_{\beta}(z, \vec x)$,
periodic in imaginary direction with the period $\beta$. Within each strip
$[\tau, \tau + \beta]$, ${\cal G}$ has two representations, one through
$G_{\beta}^+$, another through $G_{\beta}^-$, but with identical values.
This realization within one strip is repeated within all other strips.
Along the real axis, for $t > |\vec x|$, ${\cal G}_{\beta}$ has a branch
cut. The jump accross the cut is given by the canonical commutator, that is,
by the Pauli-Jordan Green function at the zero temperature. Due to the
periodicity in $\tau$ this branch cut is therefore present at $\tau = 0,
\pm \beta, \pm 2\beta, \pm 3\beta, ...$.

Let us now consider the restriction to the imaginary axis, the so-called
thermal Euclidean function, or thermal Schwinger function,
$$
{\cal G}_{\beta}(z, \vec x) \rightarrow_{t \rightarrow 0} G_{\beta}^E
(\tau, \vec x) =
\sum_n g(\vec x)
\left [ {e^{\omega_n \tau} \over e^{\beta \omega_n} - 1} +
{e^{- \omega_n \beta} \over 1 - e^{- \omega_n \beta}} \right ]~~.
$$

In zero temperature case $G_E$ was an even function, $G_E(-\tau) =
G_E(\tau)$. Here we find the same result:

{\it Problem 16.} Show that $G_{\beta}^E (-\tau) = G_{\beta}^E(\tau + \beta)
= G_{\beta}^E (\tau)$.

{\it Problem 17.} Show that
those two properties, periodicity in $\tau$ and even nature of $G_{\beta}^E$,
combined together make this function periodic with a period $\beta/2$!

As people say, $G_{\beta}^E$ is reflected around $\tau =n \beta/2$. This is
then repeated through all the other $\beta$-wide stripes.

{\it Problem 18.} Draw the analytic structure of thermal Green function
${\cal G}_{\beta}(z)$ in the complex plane $z = t + i\tau$.

\vskip0.5truein

{\bf Rindler spacetime}

\bigskip
Let us now consider the case of a flat spacetime as seen by an uniformly
accelerated observer. The trajectory of such an observer moving in
$X$ direction is given by,

$$
t= A^{-1} \sinh A T ~~,~~~~~X=A^{-1} \cosh A T ~~.
\eqno (6)$$
Here $A \in [0, \infty]$ is constant proper acceleration, and
$T \in
(-\infty, \infty)$
is his proper time. For future convenience let us introduce the dimensionless
affine parameter $\lambda \equiv A T$, and the ``radius'' $a \equiv A^{-1}$.

If we consider the set of all the observers with all possible values of
constant proper acceleration $A$ we see that pair $(\lambda, a)$ may be used
as a new set of coordinates in place of $(t, X)$.
The line element may
be written as,

$$
ds^2 = - a^2d\lambda^2 + da^2 + dl_2^2~~,
$$
where the last term stays for the length in two unchanged coordinates.
As it is well known, due to limitations on the speed to which observer may
accelerate to, these coordinates cover just the two wedges in the $(t, X)$
plane, within the straight lines $X=\pm t$. But within these wedges the
spacetime looks flat in new coordinates
$(\lambda, a)$.

Consider now the {\bf thermal} expectation value as measured by such an
observer:
${\cal G}_{\beta}^R(\lambda, a)$. Without any calculation, we know from the
preceding discussion the analytic structure of this function, in particular
we know that it must be periodic in imaginary proper time $S:~T \rightarrow S$
with a period $\beta$.

Let us compare this with the {\bf vacuum} expectation value measured by the
inertial observer using Minkowski coordinates $(t, X)$. This would
allow us to compare the two physical pictures as seen by the two observers.
The punch line is, of course, that those two pictures correspond to the
same physical situation, as we shall see shortly.
To facilitate such comparaison we only need to do a very innocent step,
just to perform the change of coordinates in one function to those naturally
used for another.
We must also take care of the proper dimensions, as it will become apparent.

We will start with the vacuum expectation value as calculated by the Minkowski
observer. Given two points, $(t_1, X_1, Y_1, Z_1)$ and $(t_2, X_2, Y_2, Z_2)$,
the vacuum expectation values must depend only on $\Delta s^2 =
-\Delta t^2 + \Delta X^2 + \Delta l_2^2$, due to the Lorentz invariance.

After substitution of the Rindler coordinates, viz. Eq. (6), we obtain
function that depends only on a finite line element expressed as,
$a_1^2 + a_2^2 - 2 a_1a_2 \cosh (\lambda_1 - \lambda_2) + \Delta l_2^2$.

One can
now make a series of simple observations.

First, note that upon change to the purely imaginary Rindler time,
$T \rightarrow iS$, equivalently, $\lambda \rightarrow i\sigma$,
the line element becomes,
$ds^2 = da^2 + a^2 d\sigma^2$. This looks like the line element in polar
coordinates, and there will be no singularity as $a \rightarrow 0$
if $\sigma$ is indeed a polar angle. But in that case it must be periodic, with
a period $2\pi$. The corresponding imaginary proper time $S$ must then also
be periodic, with a period $2\pi/A$. That is, we use $\lambda \equiv A T$
generalizes to $\lambda + i\sigma \equiv A (T + iS)$, so that
$\sigma \equiv A S$. In that case every function
of $S$ must also have the same period, or fraction of it. In particular,
we can see that for the vacuum expectation value evaluated above the Grand
vacuum Green function function {\cal G} has the form,
$$
{\cal G}(a_1^2 + a_2^2 - 2a_1 a_2 \cos(\sigma_1 - \sigma_2) + dl_2^2) ~~,
$$
along the imaginary axis. This is
the same periodicity as the thermal Green function
in flat spacetime at finite temperature $A/(2\pi)$!

This type of the reasoning is used very often: periodicity of the metric in
imaginary time is interpreted as the presence of thermal features.
However, to make reliable statement one has
to investigate the analytic structure of Green functions.

The elementary observation is that ${\cal G}(t, X)$, being the vacuum
expectation value, is holomorphic within the strip $\Delta t < \Delta X$,
i.e., for $\Delta s^2 > 0$. But this line element is invariant under the
change of coordinates, hence, this expectation value, as seen by the observer
using Rindler coordinates $(\lambda, a)$ has the same holomorphic strip and
the same periodicity in imaginary direction as what he would call the
thermal expectation value.

As a final confidence test, one can that the following statement holds:

{\it Problem 19.} Show by a direct coordinate substitution from (6) that
the equation for the Euclidean Green function in Minkowski space
becomes the equation for Euclidean Green function in Rindler coordinates.

To summarize: After transformation to Rindler coordinates,
${\cal G}(t + i\tau, \Delta x)$ has the same periodicity, and
the same analyticity strip as ${\cal G}_{\beta}(T + iS, \Delta x)$ for
$\beta =2\pi/A$; Euclidean Green functions in Minkowski and Rindler coordinates
are solutions of the two equations related by the simple change of coordinates;
the same must be said about the boundary conditions they obey; thus, it is
the same Green function; analytic continuation of Euclidean Green function
from one streep to another along the imaginary axis uniquelly defines
${\cal G}$ over the whole complex plane; therefore,
$$
{\cal G}(t + i\tau, \Delta x) = {\cal G}_{2\pi/A}(T + iS, \Delta x),
\eqno (7)$$
in the entire complex plane.

On basis of that statement we conclude that while observer using Minkowski
coordinates thinks all the time that he is computing averages with respect
to the vacuum state, the Minkowski vacuum, observer using Rindler coordinates,
the accelerated observer, notices that expectation values for a system (the
scalar field) in the same state look like thermal expectation values, computed
for a finite temperature field theory at temperature $A/(2\pi)$. In other
words, Rindler observer sees Minkowski vacuum as thermal density matrix,
temperature being proportional to its proper acceleration.

This is the main result that we wish to demonstrate here. Let us just mention
that an apparent puzzle of seeing pure state as thermal may be explained
by invoking horizons for the Rindler observer. Similarly, there is no problem
of a finite state. With an unlimited supply of energy, Rindler observer may
indefinitely maintain its constant acceleration. Once it stops accelerating,
horizons disappear, and the two observers agree that they are seeing the
vacuum state, as both are inertial observers.

\vfil\eject

\null
{\bf Schwarzschild black hole}

\bigskip
This case offers considerably more puzzle, but thanks to its similarity with
the Rindler spacetime it is easy at least to take a first step and establish
the existence of thermal features.

The standard form of the line element is,
$$
ds^2 = - \left (1 - {r_S \over r} \right ) dt^2
+ {dr^2 \over {1 - r_S/r}} + r^2d\Omega_2^2 ~~.
\eqno (8)$$
The Schwarzschild radius is $r_S = 2GM$, $M$ being the mass of the black hole.

Consider now an observer keeping himself at a fixed radius $r$ above the
horizon of the black hole. In order to do so he must accelerate away from
the hole, with an acceleration just equal to the gravitational acceleration
provided by the hole. Thus, such an observer is Rindler observer.

The line element (8) may easily be transformed into the Rindler form in
case of an observer not too far from the horizon. Let us have $r = r_S +
\delta r \equiv r_S(1 + x)$, with $x > 0$. The line element becomes
$$
ds^2 = - x dt^2 + r_S^2 {dx^2 \over x}~~.
$$
This can be transformed to the Rindler form $-a^2d\lambda^2 + da^2$ if we
define $a \equiv 2r_S x^{1/2}$, and consequently,
$$
\lambda \equiv {t \over 2r_S}~~.
$$

Upon extension to the complex plane we should find, as before, the periodicity
of the Green functions in imaginary dimensionless time $\sigma$, with period
of $2\pi$. The corresponding period in the imaginary component of the
Schwarschild time is then $\beta = 4\pi r_S$. Thus, the observer at a constant
distance above the horizon of the Schwarzschild black hole sees the
thermal bath at temperature,
$$
T_{BH} = {1 \over 4\pi r_S}~~.
$$

What is this temperature of? If the observer that maintains constant distance
from black hole is Rindler observer, then inertial observer is obviosly the
one in a free fall towards the hole. Indeed, by equivalence principle, its
vacuum state must be identical to the vacuum of the observer at infinity,
which does not detect gravitationaly the existence of the black hole. Thus,
if initial state of the quantum field is vacuum with respect to the empty
space, once black hole moves in, (or we move towards the hole), we perceive
quantum fluctuations in that state as decohered thermal fluctuations.

And why do we call this temperature the black hole temperature? One can
show, just as in the case of a Rindler observer, that horizon may be
understood as the source of the black-body radiation associated with
this temperature. By overall energy conservation this radiation must lead
to the decrease in the mass of the hole, hence to its evaporation.
\vskip0.5truein
{\bf Conclusion}
\bigskip
As said before, there are several ways to exhibit the thermal features of
black holes. There is probably no the best way, and there is no need
for one. The full understanding of the black hole evaporation most
likely will require shifting between different paradigmas and methods
of calculation. But the method reviewed here is the one that is most
firmly rooted in the quantum field theory. It clarifies what does it mean
to have finite temperature, and allows for a simple and reliable calculations.
Thus, the use of thermal Green functions ought to be usefull in future
studies of the final state of the evaporation, and the questions about
the ``loss of the quantum coherence.'' The main obstacle is how to proceed
with what is essentially a non-equilibrium configuration, and how to take
into account the back-reaction.

\vskip0.5truein
{\bf Acknowledgement}

The organizers of this Workshop must be praised for their effort and such
a splendid job under such extraordinary circumstances.

\vskip0.5truein

\def\tindent#1{\indent\llap{#1}\ignorespaces}
\def\ref{\par\hang\tindent}

{\bf References:}
\medskip
\ref{$1.$ }S.W. Hawking, {\it Nature}, {\bf 248}, 30, (1974).
\ref{$2.$ }L. Grishchuk, WUSL preprint, (1993).
\ref{$3.$ }P. Davies, S. Fulling, and W. Unruh, {\it Phys. Rev.}, {\bf D13},
2720, (1976).
\ref{$4.$ } C. Callan et al., {\it Phys. Rev.}, {\bf D45}, R1005, (1992).
\ref{$5.$ } T. Damour and R. Ruffini, {\it Phys. Rev. Lett.}, {\bf 35}, 463,
(1975).
\ref{$6.$ } G. Gibbons and M. Perry, {\it Proc. Roy. Soc.}, {\bf A 358}, 467,
(1978).
\ref{$7.$ } S. Fulling and S. Ruijsenaars, {\it Phys. Rep.}, {\bf 152}, 136,
(1987).
\ref{$8.$ } N. Bolgoliubov and D. Shirkov, {\it Vvedenije v teoriju kvantovjh
poljej}, Nauka, (1973).

\end